# Examples of usage of nearfield acoustic holography methods for far field estimations: Part 1. CW signals

*M.B. Salin, D.A. Kosteev*

*Institute of Applied Physics of the Russian Academy of Sciences, 46 Ulyanova street, 603950 Nizhny Novgorod, Russia*

A B S T R A C T

The paper is devoted to the usage of nearfield acoustic holography methods for estimating far field of the object. An experiment was carried out in anechoic chamber. First, acoustic filed was recorded in a plane that was close to source. This signals records were used to reconstruct the far field by computation routines. Second, the signal in the far field is measured and the results are compared. Several methods are tested and research on possible reduction of the microphone array size is carried out. The most significant reduction of the measurement facility complexity is usage a linear array in stead of the planar array that is made possible due to introduced computation routines.

## 1. Introduction

Carrying out acoustic measurements demands a complex organization of experiments so that their results were reliable, repeated and met requirements of metrology. Let us consider two tasks: measurement of characteristics of loudspeakers or sonars [1-3] and control measurements of machines noise levels [4]. The standard approach to noise measurement is understood as measurement of levels in far field, when it is necessary to meet the following two requirements. First, one should place the studied sound source in some area of space which can be considered as a boundless environment, or one should place the sound source into an anechoic chamber (tank). Second, one should place the microphone (hydrophone) in far zone (in Fraunhofer zone) of the sound source, so that the distance of $R$ met a condition:

$$F_r = \frac{\lambda R}{D^2} \gg 1, \qquad (1)$$

where $F_r$ is the Fresnel parameter, $\lambda$ is the sound wavelength, $D$ is the characteristic diameter of the source. It can turn out that the required distance $R$ exceeds the sizes of the available anechoic chamber. Or when the ship noise measurement is planed, since a certain distance, when the waveguide effects begin to influence, should be considered.

The following two examples show methodical errors which can arise if the measuring system is placed too close to the source and no additional data processing methods are applied. The first example is related to [1]: let a source be a dipole, and its field of pressure is given by the law:

$$p = Ae^{ikr}\left(\frac{ik}{r} - \frac{1}{r^2}\right)\cos\theta \qquad (2)$$

Here $k = 2\pi/\lambda$ is the wave number, $\theta$ is an angle between the direction of measurement and the axis of the dipole, $A$ is the complex amplitude. The signal is supposed to be harmonic, and the time dependence is $e^{-i\omega t}$. Then assume that due to absence of information about the object, one accepted a monopole model to describe the field as follows:

$$p = Aik \frac{e^{ikr}}{r}. \qquad (3)$$

The monopole model leads to the trivial ratio $|p(r_2)| = |p(r_1)| \frac{r_1}{r_2}$ which allows to recalculate results of measurement from the point $r_1$ to the point $r_2$. The possible mistake is that such calculation will be valid only for $r_1, r_2 \gg 1/k$.

The second example of a possible mistake is that various parts of a source, which is extended in space, can be coherent. The directivity pattern of radiation, measured in a far zone, will differ from the nearfield interference pattern.

However there are ways to measure an acoustic field at any distance from a sound source and find the required far field values by calculation. To do this, the amplitude and a phase of a signal has to be measured in number of points simultaneously. There is no "general direction of development" in the topic of far field prediction. Each paper can be considered as an original, trial research. The applied methods can be roughly divided into two classes. In the first one [1, 6] authors select coefficients for signals of an antenna lattice which would minimize the error for some model of a source. Sometimes derivation of calculation formulas requires regularization methods to be applied [7]. The second class of methods [2, 8, 9] is rather direct application of the Kirchhoff-Helmholtz integral. Kirchhoff-Helmholtz formula relates the distributions of pressure and its gradient measured on any closed surface with the pressure in another point outside this surface.

Such scheme of measurement is seldom implemented directly, because of its complexity. For example, gradient measurements demand devices to be calibrated and positioned with a high precision [2]. Instead of that, first of all it is possible to make the computational formulas free of the term, which contain the gradient. This can be done by selecting the proper kind of the Green's function. Second, one should try to avoid using closed measuring surfaces in order to make the measurement device more practical. Ways to calculate pressure levels and directivity patterns in far field, using the results of nearfield measurements, are proposed in [5] for the case when such measurement are done with planar or even linear array. This paper presents a rather simple experiment that has been made to check several known methods of recalculation of the neadfield data into the far filed. The algorithms given by [5] are considered among the other ones.

One should note that the first works on nearfield measurements have been performed in a radar-location them. Directivity patterns of big radar antennas were studied in this way. The main directions and results of works are given by [10] and the current state is given by [11]. The main difference of acoustic nearfield measurements from radiophysical ones is that acoustical signals are not so stationary and positions of sensors are usually given with some uncertainty. Besides that acoustic signals are usually wide band, however this is going to be studied in the follow-up paper.

The other kind of tasks is considered in nearfield acoustic holography as well. A big number of papers is devoted to estimation of pressure values in the points that are closer to the sound source than the measuring surface [12-14]. Unlike the tasks, which are related to subwavelength, the below theory does not require regularization of mathematical ratios.

## 2. Experimental setup

The experiment was carried out in an anechoic chamber. The scheme of the installation is provided on fig. 1. The installation included: a) two separated loudspeakers which imitated a sound source with non-isotropic directivity pattern and b) the linear array (antenna) of microphones, which were equidistant located at a common cable. In the first part of the experiment the microphone array was consistently moved in the plane $z=z_{near}$=const. This plane was at the small distance from loudspeakers. The measurements taken at various provisions of the array was combined together to obtain a general view of the acoustic field in the $z$=const plane. These measurements were the input data for the algorithm that calculated parameters of the acoustic field at $z=z_{far}$. In the second part of the experiment the array of microphones was set at distance $z_{far}$ to make reference measurements of the far field levels.

One loudspeaker was shifted on $D$=49 cm along the $x$ axis relative to the other one, and they transmitted antiphase continuous (tonal) signals at frequencies of 500 and 1500 Hz. Loudspeakers were used without enclosure, i.e. both sides of the membrane of the loudspeaker were open. The distance from the sources (the plane of the external part of the loudspeakers) to the array at its closest position was $z_{near}$ = 28 cm. The distance at the most remote position was $z_{far}$ = 203 cm.

The microphone array length was 200 cm (along the $x$ axis), the distance between microphones was 10 cm. The array was moved was moved within the interval of -61.5..60 cm along the $y$ axis with the average step of 12 cm, where $y$=0 is the coordinate of the loudspeakers. Thus, the spatial sampling step was less than the wavelength 1500 Hz, and less than a half of the wavelength at 500 Hz. An extra microphone, which was connected to the same milticore cable and to the same data acquisition system, stayed immobile and was used as a reference in the whole series of measurements.

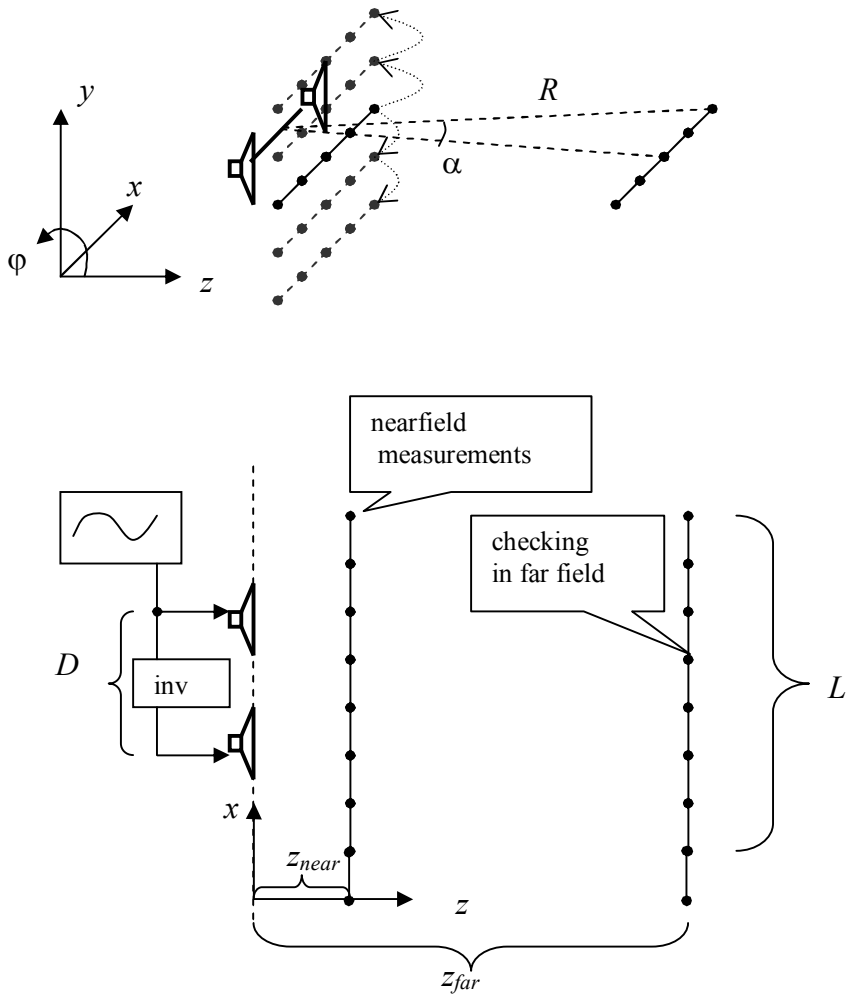

**Fig. 1.** Scheme of experimental setup. Distance values are given in the text.

The values of Fresnel parameter (1), which have been calculated according to the frequencies and the geometrical sizes, are specified in Table 1. These values show that one can truly call the closest and the remote position of the array near field and far field.

**Table 1**

Values of Fresnel parameter $F_r$.

| Frequency (Hz) | $F_r$ at $r=z_{near}$ | $F_r$ at $r=z_{far}$ |
|---|---|---|
| 500 | 0.7 | 5.1 |
| 1500 | 0.2 | 1.7 |

Initial results of measurements are represented by multichannel files, containing the signal records. Each file corresponds to one position of the array. Since the loudspeakers transmitted only a set of tones in this experiment, the received signal undergone filtering and heterodyning so that complex amplitudes at each microphone were obtained. Nearfield records for each posi-tion *y* were by complex coefficients to make both amplitude and phase of a signal of the reference microphone were constant. (Recall that the reference microphone was immobile while array was moved.) Thus, a planar array was been synthesized from a set of lines.

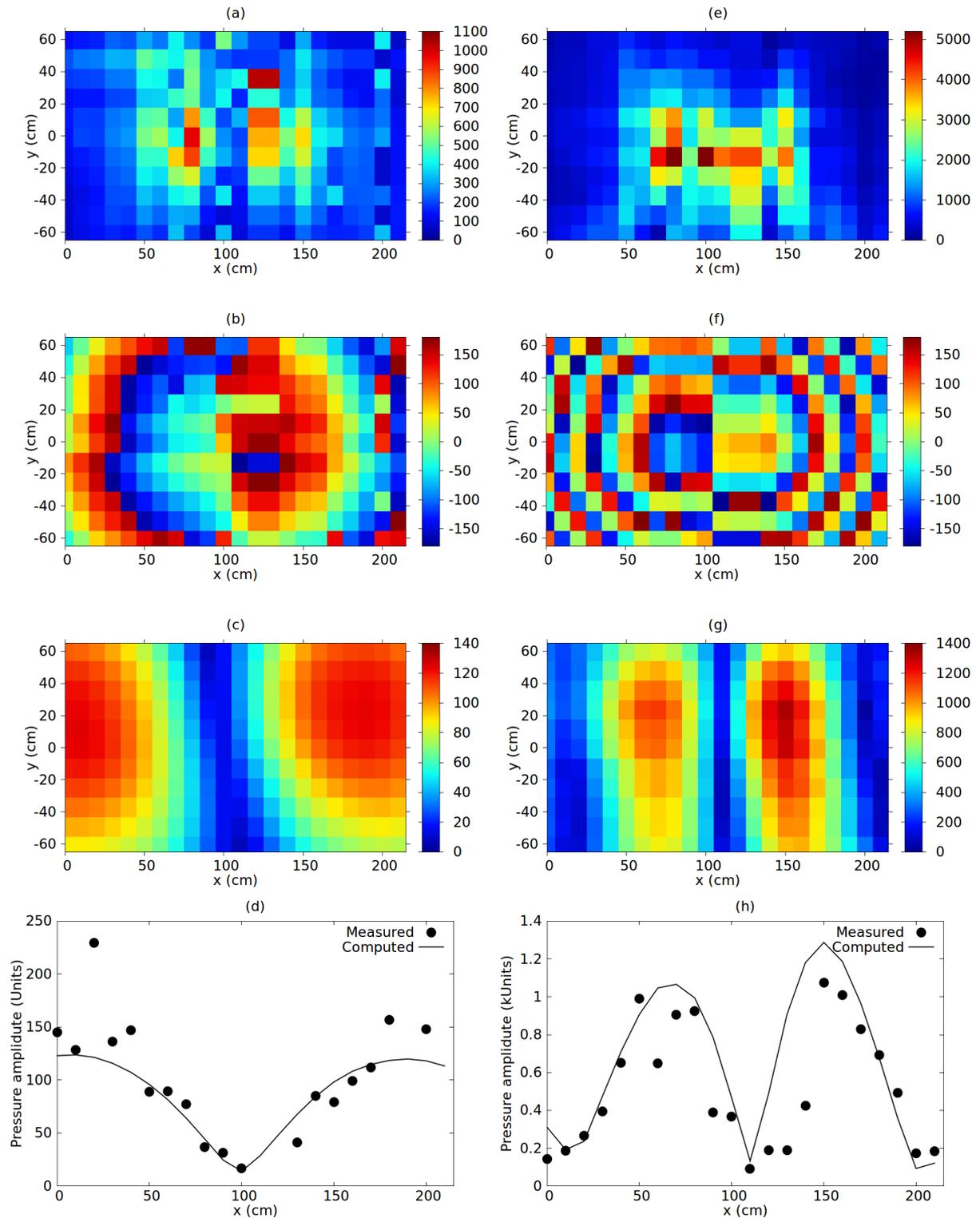

**Fig. 2**. Results of measurement and calculation of the field of a complex source: (a,e) nearfield measurement of pressure amplitude, given in conventional units; (b,f) nearfield measurement of a phase (degrees); (c,g) computed values of the amplitude in the far field (FPS method is used, see the text for details); (d,h) comparison of measured and computed values for the far field (distribution along *x* axis at *y*=0). The entire left column (a-d) corresponds to 500 Hz, the right column (e-h) corresponds to 1500 Hz.

The distribution of amplitude, measured in the $z_{near}$ plane, is plotted on fig. 2a,e and distribution of a phase is plotted in fig. 2b,f. From figure it can be seen that sources work in an antiphase. Unfortunately, the image has distortions due to defects of the measuring system. Looking ahead one can tell that fig. 2c show the far field values ($z=z_{far}$ plane), which were computed basing on fig. 2a and fig. 2b. Fig. 2g was computed basing on fig. 2e,f respectively. The computation method will be discussed below. It is clear that the intensity minimum appears in the center due to the antiphase signals were transmitted from to points. Two maxima of the far field directional pattern can be registered within the measuring aperture. The positions of the maxima depend on frequency. Fig. 2d and fig 2h show comparison of measured and computed values for the far field. The section of $y=0$, $z=z_{far}$ is plotted. (Far field measurements were taken along a single line.)

## 3. Methods of calculation of the far field

*3.1 FPS method: Far field prediction by decomposing a Plane array data into plane-wave Series*

Arbitrary distribution of pressure in half-space $z>0$ can be decomposed into plane waves as follows [8]:

$$p(x,y,z) = \sum_l \sum_m \tilde{a}_{lm} \exp(ik_{xm}x + ik_{yl}y + ik_{zml}z), \qquad (4)$$

where $p$ is a complex amplitude of pressure, $\tilde{a}_{lm}$ are coefficients, $\{k_{xm}, k_{yl}, k_{zml}\}$ is a wave vector, which corresponds to a spatial harmonica ($l, m$). $k_{zml}$ is a function of $k_{xm}$ and $k_{yl}$ defined as follows: $k_{zml} = \sqrt{\frac{\omega^2}{c^2} - k_{xm}^2 - k_{yl}^2}$, where $\omega$ is a cyclic frequency, $c$ is the sound speed, which is assumed to be 300 m/s (special measurements of $c$ were not carried out). If $\omega^2/c^2 < k_{xm}^2 + k_{yl}^2$ then the above expression for $k_{zml}$ should be considered as follows: $k_{zml} = i\sqrt{k_{xm}^2 + k_{yl}^2 - \omega^2/c^2}$. The time function $e^{-i\omega t}$ is omitted. The sign of $k_{zml}$ is chosen considering the fact that sources are located in the $z \le 0$ half-space.

Values of complex amplitude of $p(x_n, y_n, z_{near})$ that are measured in nodes of a square grid $x_n, y_n$ in the nearfield can be used for far field estimation via the following expressions:

$$a_{lm} = \frac{1}{S} \sum_{j=-J/2}^{J/2} \sum_{n=-N/2}^{N/2} \Delta s_{jn} h(x_n, y_j) p(x_n, y_j, z_{near}) \exp(-ik_{xm}x_m - ik_{yl}y_j) \qquad (5)$$

$$p(x, y, z_{far}) = \frac{JN}{M^2} \sum_{l=-M/2}^{M/2} \sum_{m=-M/2}^{M/2} a_{lm} \exp\left[ik_{xm}x + ik_{yl}y + ik_{zml}(z_{far} - z_{near})\right] \qquad (6)$$

where $a_{lm}$ is an estimated value of $\tilde{a}_{lm}$ ($a_{lm}$ can differ from $\tilde{a}_{lm}$ due to introduced coefficients), $N$ and $J$ are the number of elements of the planar array by $x$ and $y$ axis respectively (number of microphones in a line and number of shifts of the entire line), $M$ is the number of harmonics used for reconstruction of the field (the optimal value of $M$ will be discussed below), $\Delta s_{jn}$ is the area of an element, $h$ is the Hann window for a two-dimensional case:

$$h(x_n, y_j) = \frac{1}{4}\left(1 - \cos\frac{2\pi(x_n - x_1 + \Delta x/2)}{x_N - x_1 + \Delta x}\right)\left(1 - \cos\frac{2\pi(y_j - y_1 + \Delta \bar{y}/2)}{y_N - y_1 + \Delta \bar{y}}\right). \tag{7}$$

Here $\Delta x$ is the interelement distance over $x$, $\Delta \bar{y}$ is the average step over $y$. The area of an element is calculated as follows, except for the case of extreme values of $j$ and $n$.

$$\Delta s_{jn} = (x_{n+1} - x_{n-1})(y_{j+1} - y_{j-1})/4, \tag{8}$$

The following expression is used for the left edge that is $n=1$:

$$\Delta s_{jn} = (x_n - x_{n-1})(y_{j+1} - y_{j-1})/4, \tag{9}$$

and other extreme points are treated in the same fashion.

Pay attention that the forward Fourier transform is used in such way that it produces greater number of output values than the number of input values that is given. Namely the input values for Eq. (5) are $NJ$ signal samples, i.e. all available measuring points. The output values are $M^2$ spatial harmonics with $M>N$ and $M>J$. Such number of harmonics is used in Eq. (6). This was done like expanding the field of measurement by adding zero values. This helps to avoid a problem of the following kind. The usage of the Fourier transform explicitly governs that the problem is periodical on $x$ and $y$, and additional imaginary sources may appear. For example, point (0, 0, 0) is copied to ($M\Delta x$, 0, 0), (0, $M\Delta \bar{y}$, 0), (-$M\Delta x$, 0, 0) and so on. The idea is to move this imaginary sources to such large distance so they will give almost no contribution at $z=z_{far}$. As a result, $M$ has to be chosen from next criterion:

$$M \gg (z_{far} - z_{near})/\Delta x, \tag{10}$$

$$M \gg (z_{far} - z_{near})/\Delta \bar{y}. \tag{11}$$

In the below calculations $M$ was set equal to 200.

The sizes of the sound source are required to be less than the sizes of the array in order to apply the discussed method. Section 4 describes an experimental study on how large should be the difference of the sizes. As for theory, according to [5] the antenna has to be prolonged beyond the source dimensions at least on the size of the first Fresnel zone that is $\sqrt{\lambda x_{near}}$. In other words, the most part of the energy flux, emitted to the interesting point, should pass through an aperture of the array. This formulation explains why the infinite plane can be replaced by an an-tenna of a finite area. The aperture size in the presented experiment is sufficient. Here we lose some information about the waves, which propagate under big angles to axis *z*. But these waves don't make a great contribution in that area of space which is studied in the far field.

So the results of the calculation of the far field by this method have been shown above in fig. 2. Fig.2dh shows that the results of the benchmark measurement, which was held in a far zone, coincide with the calculated results rather well.

This calculation method will be refered furthered as FPS. The method is rather general. There is no strict requirement that $z_{far}$ belongs to a far zone (Fraunhofer zone). The method of re-calculation is valid for the Fresnel zone too.

*3.2 FPK method: Far field prediction from a Planar array with use of the Kirchhoff integral*

Is sources are concluded inside the closed surface *S* then their field at in any outside point (in free space) can be computed by means of Kirchhoff-Helmholtz integral [15, 16]:

$$p(\mathbf{r}') = \int_S \left( p(\mathbf{r}) \frac{\partial G(\mathbf{r},\mathbf{r}')}{\partial n} - \frac{\partial p(\mathbf{r})}{\partial n} G(\mathbf{r},\mathbf{r}') \right) dS \qquad (12)$$

where **r** lies on surface *S*, **r'** is a point outside of *S*, *n* is an external normal to *S*, *G* is the Green's function. In relation to the task solved here, the $z=z_{near}$ plane can be considered as *S*. First, one may consider infinite plane as a closed surface due to it contains the infinitely far point. Second, proceeding from the aforesaid in section 3.1 the infinite integral is replaced by the integral over the finite area of the existing array. Note that a similar approach was used in [2].

Consider additional transformations of Eg. (12) following [5, 16]. Taking into account the fact that the pressure gradient was not measured in experiment, the so-called Green's function of a soft body should be used:

$$G(\mathbf{r},\mathbf{r}') = \frac{1}{4\pi} \left( \frac{\exp(ikR')}{R'} - \frac{\exp(ikR'')}{R''} \right) \qquad (13)$$

where **r**={*x, y, z*}, **r'**={*x', y', $z_{far}$*},

$$R' = \sqrt{(x-x')^2 + (y-y')^2 + (z-z_{far})^2}$$

$$R'' = \sqrt{(x-x')^2 + (y-y')^2 + (z-2z_{near}+z_{far})^2}.$$

After substitution of (13) in (12), one can notice that the term containing pressure gradient becomes identically equal to zero. Now require that $z_{far}$ in strictly in the far field, i.e. inequality (1) is satisfied with the "much more" sign. Then spherical waves in expression (13) can be replaced by plane waves. Finally we come to the following equation:

$$p(x,y,z_{far}) = -\frac{ik\cos\alpha}{2\pi R} \sum_{j=-J/2}^{J/2} \sum_{n=-N/2}^{N/2} \Delta s_{jn} h(x_n, y_j) p(x_n, y_j, z_{near}) \exp\left(-ik\frac{x}{R}x_n - ik\frac{y}{R}y_j\right) \quad (14)$$

where $R$ is a distance from the center of the array to a point ($x$, $y$, $z_{far}$), $\alpha$ is an angle between a normal to the array and the direction to the point ($x$, $y$, $z_{far}$), so that $\cos\alpha = (z_{far} - z_{near})/R$.

Results of calculation of the far field by the described method are constructed in fig. 3. Actually, the results of computation by all method that are discussed in the paper are collected in the mentioned fig. 3. The results of the benchmark measurement are present in fig. 3 as well. The method described in this section is designated in fig. 3 as FPK. The divergence of 27% between the results of calculation and the experiment is observed for the frequency of 1500 Hz. It can be explained by the fact that the Fresnel parameter is not big enough at this frequency (see table 1).

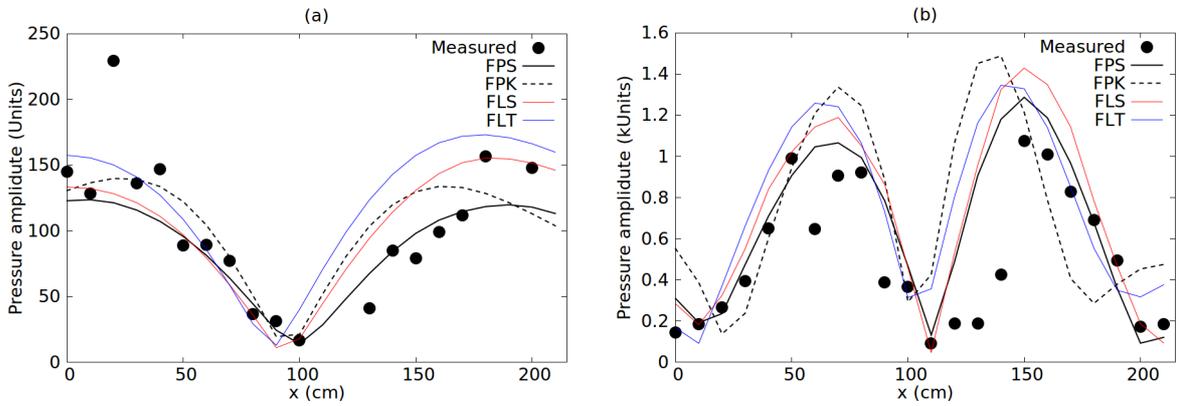

**Fig. 3.** Comparison of results of the measurement and the calculations, done by various methods. Distribution of pressure amplitudes in a far flied along the line $y=0$, $z=z_{far}$, the frequency is: a) 500 Hz, b) 1500 Hz.

*3.3 FLS method: Far field prediction by decomposing a Linear array data by decomposition into cylindrical wave Series*

As it was mentioned above, a certain practical interest is connected with a possibility to reduce a number of sensors in a measuring system. Salin [17] proved the possibility to apply near-field holography methods to three-dimensional objects when carrying out measurements by a one-dimensional (linear) array. Limitations on when this can be done were discussed in [17] as well. Similar ratios are derived in this paper but in a different way. Regarding the conditions of the experiment, which is discussed in this paper, we are going to use only data, recorded at single position of the array, namely $y=0$.

Consider the cylindrical system of coordinates with axis $x$, angle $\varphi$ (as shown in fig. 1) and radial coordinate $\rho = \sqrt{y^2 + z^2}$. The acoustic field can be decomposed into cylindrical waves as follows:

$$p(x,\varphi,\rho) = \sum_l \sum_m \left[ \tilde{b}_{lm} H_l(\kappa_m \rho)\cos(l\varphi) + \tilde{c}_{lm} H_l(\kappa_m \rho)\sin(l\varphi) \right] \exp(ik_{xm}x) \quad (15)$$

where $\tilde{b}_{lm}$, $\tilde{c}_{lm}$ are coefficients, $H_l$ is the Hankel function of order $l$ and of the 1st kind (outgoing wave if the time function is $e^{-i\omega t}$). Wave number $\kappa_m$ is defined as follows: $\kappa_m = \sqrt{k^2 - k_{xm}^2}$, if $k^2 \geq k_x^2$, otherwise $\kappa_m = i\sqrt{k_{xm}^2 - k^2}$. In the second case $H_l(\kappa_m \rho)$ fades, when $\rho$ increases.

The main idea of the proposed method is to eliminate summation over $l$ in Eq. (15). To do this one should specify the following limitation of the measurement method.

1. The source has to be extended along the axis, which is parallel to an axis of the linear array, (along axis $x$ in case of Fig. 1).
2. The array has to longer that the sound source at a size that will be concretized below, in section 4.
3. One is going to calculate the field in a half-plane formed by the source axis and the array axis (the half-plane of $y=y_0=0$ and $z>0$ in case of Fig.1, that is $\varphi=0$ half plane).
4. The angle, where the reconstructed field is valid, is limited by introducing a threshold like $k_{xm} \leq k_{x,\max}$. It means that waves, extending under small corners to an axis $x$ are excluded from consideration.
5. The source should not have too complicated directional pattern in the plane $yz$.

Rewrite issue 4 so that the decomposition, given by (15), should be limited according to azimuthal indexes by $l \leq l_{max}$. Recall that in case of $\kappa\rho \gg \sqrt{l+1}$ the following expression is valid:

$$H_l(\kappa\rho) \approx \sqrt{\frac{2}{\pi\kappa\rho}} \exp\left[i\left(\kappa\rho - \frac{\pi l}{2} - \frac{\pi}{4}\right)\right] \quad (16)$$

Combination of the above mentioned criteria leads to the following equation:

$$k_{xm} \leq k_{x,\max} = \sqrt{k^2 + \frac{(l_{\max}+1)}{z_{near}^2}} \quad (17)$$

The patter of the source radiation over indexes $l$ is connected to its wave sizes in cross section. The previous works, including [17], showed that in most cases (17) may be replaced by a simpler rule:

$$\frac{\lambda z_{near}}{D_{yz}^2} \gg 1 \quad (18)$$

Expression (18) is similar to the parameter (1), but $D_{yz}$ is used herm which is the characteristic size of a body in the cross section of $yz$.

Substituting (16) in (15) and putting $\varphi=0$, one can rewrite (15) as:

$$p(x, y_0, z) = \sum_m \tilde{b}_m \sqrt{\frac{2}{\pi \kappa_m z}} \exp\left[i\kappa_m z - i\frac{\pi}{4}\right] \exp(ik_{xm} x) \qquad (19)$$

where $\tilde{b}_m = \sum_l \tilde{b}_{lm} \exp(-i\pi l/2)$ which is a set of all coefficients with different $l$, combined in one general coefficient. Thus, if the field of an extended source is considered in a half-plane and the condition (17) or (18) is satisfied then the equation, which describes the cylindrical wave expansion, does not depend on azimuthal structure of the received field.

On the basis of (19) the following working formulas can be derived for recalculating near field into the far field:

$$b_m = \frac{1}{N} \sum_{n=-N/2}^{N/2} h_1(x_n) p(x_n, y_0, z_{near}) \exp(-ik_{xm} x_n) \qquad (20)$$

$$p(x, y_0, z_{far}) = \frac{N}{M} \sqrt{\frac{z_{near}}{z_{far}}} \sum_{m=-M/2}^{M/2} b_m F(\kappa) \exp\left[ik_{xm} x + ik_{zml}(z_{far} - z_{near})\right] \qquad (21)$$

where $b_m$ are the estimations if $\tilde{b}_m$ (that may differ from $\tilde{b}_m$ due to the introduced coefficients), $F(\kappa)$ is a filter suppressing harmonics which do not satisfy (17) and $h_1$ is the Hann function:

$$h_1(x_n) = \frac{1}{W}\left(1 - \cos\frac{2\pi(x_n - x_1 + \Delta x/2)}{x_N - x_1 + \Delta x}\right) \qquad (22)$$

For this case it was found that the best results are obtained by choosing such a normalizing coefficient $W$, so that $\sum_{n=1}^{N} h_1^2(x_n) = N$.

Results of calculation of the far field which was carried out by means of the method considered in this section are plotted in fig. 3, and the line corresponding to them is designated by FLS. A divergence with reference measurement of the far field is greater than for the above methods where the planar antenna was exploited. However the FLS method shows acceptable quality of calculation too.

Note that it is indeed possible to restore the far field of a three-dimensional object by measuring in the near zone with a single linear array. Recall that the sound source in this experiment was a set of loudspeakers without enclosure, so a directional pattern of each of them was a dipole (in the $yz$ plane). When processing signals, this information was not used in any way, and the measurement with a line of microphones, distributed along the $x$ axis does not allow determining this diagram in principle. Nevertheless, the far filed calculation was rather correct.

## 3.4 FLT method: Far field prediction from a Linear array processing via the Transfer function

The last method considered in this article will be denoted FLT. The method is reproduced according to [5]. The method also deals with the linear array processing and the limitations are the same as mentioned above, in section 3.3. And the requirement that $z_{far}$ is strictly in the far field is added here. Ommiting the derivation, given in [5], the final equation to calculate the filed in a point (R$\sin\alpha$, $y_0$, R$\cos\alpha$) may be written as follows:

$$|p(\alpha,R)| = \frac{|b_{m*}|N\Delta x}{R}\sqrt{\frac{\omega x_{near}\cos\alpha}{2\pi c}}, \qquad (23)$$

where $b_{m*}$ is given by Eq. (20) and it is a function of the nearfield measured data. Index $m*$ is chosen so that $k_{xm*}/k = \sin\alpha$. Results of calculation by this method are plotted in fig. 3 (see curve FLT). They were close to results of the previous method.

Note that the signal amplitude in a far zone is proportional to the output of the array phased on the same angle of $\alpha$ and located in a near zone. This follows from (20) and (23). The same feature might have been mentioned above for the FPK method, see Eq. (14).

## 4.1 Influence of the array length on results of the far field reconstruction

It was said above that the most part of the energy flux, emitted to the interesting point, should pass through an aperture of the array. This requirement belongs to all four considered methods. Let us consider the FLS method as a basis and investigate an error of the far field reconstruction depending on the array length. 22-element line array of microphones was used in the experiment. Subsets with smaller number of channels are going to be selected out of the recorded multichannel files. The subsets will undergo the same processing as it was done for the full length array in section 3.3.

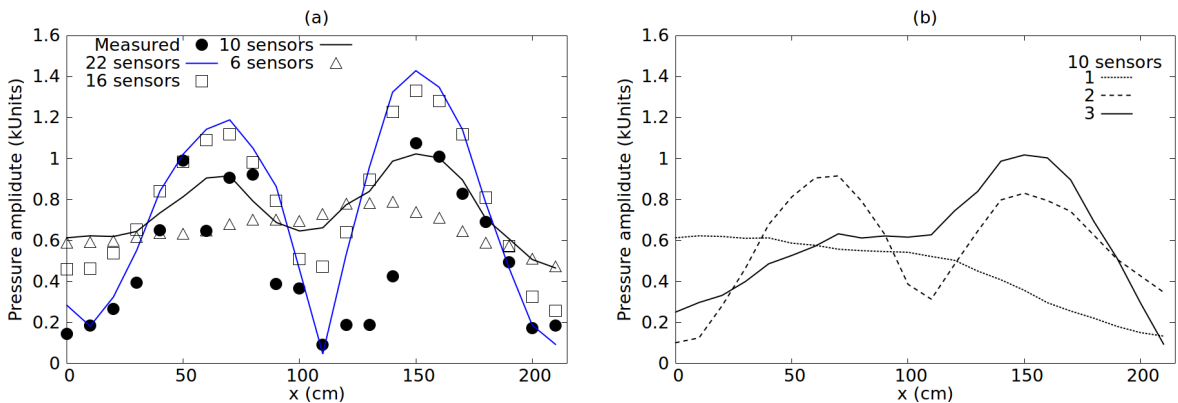

**Fig. 4.** Results of measurements by arrays of various length and application of the FLS method for calculation of the far field: a) the "hold max" procedure is applied, b) the distributions of the field, calculated for various positions of the array prior to the "hold max" procedure. Frequency 1500 Hz.

Fig. 4b shows the result of the far field calculation, using a subset of 10 microphones. It is possible to begin selecting the subset stating with a different position of the microphone. Three curves are plotted for different starting points. The obtained subsets are shifted on 30 cm from each other. Fig. 4b shows that the obtained far field patterns vary. Recall that the sound source creates two (at least two) lobes of the pattern. If we take the most right subset of microphones for processing, then the curve 3 is obtained, which highlight the right lobe. The same is valid for the left one.

The "hold max" procedure is proposed to obtain the correct directional pattern in case of short receiving array. Consider, for instance, the results of measurements $q = 1, 2$ and $3$ in fig. 4b. Let $p_q(x_n, y_0, z_{far})$ be a result of processing of each $q$-th position of the array. Create a table of values $p_{max}(x_n, y_0, z_{far})$ as a function of $n$. For each $n$ execute operation:

$$p_{max}(x_n, y_0, z_{far}) = \max_q |p_q(x_n, y_0, z_{far})|. \tag{24}$$

The obtained $p_{max}(x_n, y_0, z_{far})$ is the requested distribution of amplitude of the field.

Finally, several sizes of subsets were considered, and the result of the far field prediction are plotted on Fig. 4a. The "hold max" procedure has been already applied there. The plot shows that 10 is the minimum number of sensors that gives satisfactory results. Such subset has a length of 90 cm that is approximately the length of the source plus two radii of the 1st Fresnel zone. (Recall that the source length is $D=49$ cm and the radius of the 1st Fresnel zone is $\sqrt{\lambda x_{near}} = 23.7$ cm at the frequency of 1500 Hz). The subset of 6 sensors does not reproduce the two lobes and the zero value in the center.

The experiment models a situation which took place in [18] where a research of an underwater sound source was carried out. A linear array, that was not long enough, stayed stationary while a sound source was moving along it. Fragments of a signal record in various time moments corresponded to a various positions of the source relative to the array. Various directional patterns were obtained from the different fragments of the signal, so the FLT method, explained in section 3.4, was applied, followed by the "hold max" routine, given by Eq. (24). In a sea experiment it is rather difficult to operate extended arrays of hydrophones, so the possibility of work with less bulky equipment is always welcomed.

**Acknowledgements**

This work was funded by the Russian program of the basic scientific research of the state academies of sciences (ref. no. 12.18) and Russian-Chinese joint laboratory "One belt, one road" (established by Zhejiang province of China). The authors would like to acknowledge the contribution of Huancai Lu and Zubin Liu of Zhejiang University of technology, and we hope that they will be coauthors of the next chapter of the paper.